\documentstyle[11pt]{article}

\font\mbld=cmmib10 scaled \magstep1
\def\ttt{\mbld \char'034}

\begin{document}

\date{}

\begin{titlepage}

\title{On multi-particle functions and pion-decay constant
in Nambu--Jona-Lasinio model}

\author{
 V E Rochev\footnote{Institute for High Energy Physics,
 Protvino, Moscow region, Russia
 (e-mail: rochev@ihep.ru)} }
\date{}

\end{titlepage}
\maketitle
\begin{abstract}
The system of  second-order equations of  mean-field expansion is
considered  for  Nambu--Jona-Lasinio model with chiral symmetry of
SU(2)-group. The system includes equations for four-particle and
three-particle functions and also equations for next-to-leading
two-particle function and next-to-next-to-leading quark
propagator. Exact solutions for four-particle and three-particle
equations are obtained. The solution for four-particle function is
a disconnected combination of the leading-order two-particle
functions. The connected part of three-particle function includes
two-meson and three-meson contributions. The solution of
three-particle equation permits to close the equation for
next-to-leading two-particle function and to calculate a
correction to pion-decay constant $f_\pi$. This correction
increases from 14\% to 28\% when values of quark condensate varies
from $-(210 MeV)^3$ to $-(250 MeV)^3$.

\end{abstract}

\newcommand{\ba}{\begin{eqnarray}}
\newcommand{\ea}{\end{eqnarray}}
\newcommand{\tr}{\,\mbox{tr}\,}

\section*{Introduction and Summary}
Nambu-Jona-Lasinio (NJL) model with quark content is one of the
most successful effective models of quantum chromodynamics of
light hadrons (for  review see \cite{kle}-\cite{HatKun}).  A
number of
 physical applications of NJL model is connected with
multi-quark functions, which are the subject of present work.
These multi-quark functions arise in higher orders of the
mean-field expansion (MFE) for NJL model. To formulate MFE we have
used an iteration scheme of solution of Schwinger-Dyson equation
with fermion bilocal source, which has been developed in works
\cite{Ro1}. We have considered equations for Green functions of
NJL model up to the second order of MFE. The leading approximation
and the first order of MFE maintain equations for the quark
propagator and the two-particle function and also the
next-to-leading (NLO) correction to the quark propagator. The
second order of MFE includes the equations for four-particle and
three-particle functions and also equations for the NLO
two-particle function and the next-to-next-to-leading (NNLO) quark
propagator.

We have found  solutions for the  four-particle and three-particle
equations. The solution for four-particle function is a
disconnected combination of the leading-order two-particle
functions, consequently,  the physical effects, which  connected
with four-particle function (i.e., pion-pion scattering),  are
suppressed in this order of MFE.

The solution of four-particle equation gives us a possibility to
close the equation for  three-particle function. The solution of
three-particle equation contains both disconnected and connected
parts. The connected part of three-particle function includes
two-meson and three-meson contributions.

 The solution of
three-particle equation permits to close the equation for NLO
two-particle function and to calculate a correction to pion-decay
constant $f_\pi$. This correction varies from 14\% to 28\% when
values of quark condensate varying from $-(210 MeV)^3$ to $-(250
MeV)^3$.

The correction to pion-decay constant has been calculated early in
work \cite{Nikolov}  in the framework of improved $1/n_c$
expansion by using the partially bosonized version of NJL model
(see also \cite{Oertel}). Our result  is in accordance with
results of these work, though the method of the calculation is
 different. This fact demonstrates a self-consistency of NJL model
concerning to high-order corrections.

\section{Mean-field expansion in bilocal-source formalism}
We consider NJL model with chiral symmetry $SU_V(2)\times
SU_A(2)$.  The model Lagrangian is
\begin{equation}
{\cal L}=\bar \psi i\hat \partial\psi+\frac{g}{2}
\biggl[(\bar\psi\psi)^2+(\bar\psi i\gamma_5
\mbox{\ttt}\psi)^2\biggr]. \label{LSU2}
\end{equation}
Here $ \psi$ is the quark field with  $n_c$ colours, $g$ is the
coupling constant of  $m^{-2}$ dimension, $\mbox{\ttt}$ are Pauli
matrices.

The mean-field expansion in the bilocal-source formalism for the
model can be constructed with the method of work  \cite{Ro1} (see
also \cite{JaRo}).

Generating functional $G$ of Green functions is the functional
integral
\begin{equation}
G(\eta)  =\int D(\psi,\bar\psi)\exp i\Big\{\int dx{\cal L} -\int
dx dy \bar\psi(y)\eta(y,x)\psi(x)\Big\}. \label{G}
\end{equation}
Here $\eta(y,x)$ is the bilocal source of quark field. $n$--th
derivative of $G$ over source   $\eta$ is  $n$-particle
($2n$--point) Green function  $ S_n\left(
\begin{array}{cc} x_1&y_1\\\cdots&\cdots\\x_n&y_n\end{array}
\right).$

As a consequence of translational invariance of the
functional-integration measure in (\ref{G}) we have the
functional-derivative Schwinger-Dyson equation:
\begin{equation}
\delta(x-y)G + i\hat\partial_x\frac{\delta G}{\delta\eta(y,x)}
+ig\Big\{\frac{\delta}{\delta\eta(y,x)}\tr\Big[\frac{\delta
G}{\delta\eta(x,x)}\Big]- \gamma_5 \mbox{\ttt}
\frac{\delta}{\delta\eta(y,x)}\tr\Big[\gamma_5
\mbox{\ttt}\frac{\delta G}{\delta\eta(x,x)} \Big]\Big\} =
\label{SDE}
\end{equation}
$$
=\int dx_1\eta(x,x_1) \frac{\delta G}{\delta\eta(y,x_1)}.
$$
As a leading approximation for the mean-field expansion we
consider equation (\ref{SDE}) without right-hand-side. A solution
of the leading-order equation is the functional
\begin{equation}
G^{(0)}= \exp\Big\{\mbox{Tr}\,\Big[S\ast\eta\Big]\Big\}
\label{G^0}
\end{equation}
(Here and below $\mbox{Tr}$ denotes the operator trace, and $\ast$
denotes the operator multiplication.) Function $S$ is a solution
of the equation
\begin{equation}
\delta(x)+i\hat\partial S(x)  +ig S(x)\tr [S(0)]=0. \label{S0}
\end{equation}
This equation is in essence the coordinate form of the well-known
gap equation for the quark propagator.  In the leading
approximation the unique connected Green function is quark
propagator
$$
 S_1^{(0)}\equiv S = (m-\hat
p)^{-1}.
$$
Here $m$ is a quark dynamical mass.

 The leading approximation (\ref{G^0}) generates the linear
 iteration scheme:
$$
G = G^{(0)} + G^{(1)} + \cdots + G^{(n)} + \cdots,
$$
where functional   $G^{(n)}$ is a solution of the iteration-scheme
equation
\begin{equation}
G^{(n)} + i\hat\partial \frac{\delta G^{(n)}}{\delta\eta}
+ig\Big\{\frac{\delta}{\delta\eta}\tr\Big[\frac{\delta
G^{(n)}}{\delta\eta}\Big]-
\gamma_5\mbox{\ttt}\frac{\delta}{\delta\eta}
\tr\Big[\gamma_5\mbox{\ttt}\frac{\delta
G^{(n)}}{\delta\eta}\Big]\Big\} =\eta\ast \frac{\delta
G^{(n-1)}}{\delta\eta}. \label{Gn}
\end{equation}
The general solution of equation (\ref{Gn}) is the functional
$$
G^{(n)}= P^{(n)}G^{(0)},
$$
where  $P^{(n)}$ is a polynomial of  $2n$-th degree on source
$\eta$.
\\
\section{First-step equations }
The functional of the first step of the iteration scheme is
$$
\biggl\{\frac{1}{2}\mbox{Tr}\Big[ S_2\ast\eta^2\Big] +
\mbox{Tr}\Big[S^{(1)}\ast\eta\Big]\biggr\}G^{(0)}.
$$
Here $S_2$ is the two-particle function, $S^{(1)}$ is the NLO
quark propagator. Taking into account
 the  leading-order equations we have the system of equations for
 $S_2$ and  $S^{(1)}$.

The equation for $S_2$ has the form
\begin{equation}
S_2\left( \begin{array}{cc} x&y\\x'&y'\end{array} \right)= -
S(x-y') S(x'-y)+\label{S2}
\end{equation}
$$
 +ig\int dx_1\Big\{(S(x-x_1)
S(x_1-y)) \tr_u\Big[ S_2\left( \begin{array}{cc}
x_1&x_1\\x'&y'\end{array} \right)\Big]-
$$
$$
 - (S(x-x_1)\gamma_5\mbox{\ttt}
S(x_1-y)) \tr_u\Big[ \gamma_5\mbox{\ttt}S_2\left(
\begin{array}{cc} x_1&x_1\\x'&y'\end{array} \right)\Big]\Big\}
$$
Here   $\tr_u$ denotes the trace, which includes the upper line of
function $S_2$.

A solution of equation (\ref{S2}) is well-known. Usually such
equation is solved in momentum space. We give the solution of this
equation in coordinate space since a similar method will be used
for solving the equation for the three-particle function.

Let us go to the amputated function
$$
F_2=S^{-1}\ast S^{-1}\ast S_2\ast S^{-1}\ast S^{-1}
$$
and define the connected part
$$
F^c_2\left( \begin{array}{cc} x&y\\x'&y'\end{array} \right)=
F_2\left( \begin{array}{cc} x&y\\x'&y'\end{array}
\right)+S^{-1}(x-y')S^{-1}(x'-y).
$$
Then for the function  $F_2^c$ we obtain the equation
$$
F^c_2\left( \begin{array}{cc} x&y\\x'&y'\end{array} \right)=
-ig\delta(x-y)\delta(x-y')\delta(x'-y)\{1\cdot
1-\gamma_5\mbox{\ttt}\cdot\gamma_5\mbox{\ttt}\}+
$$
$$
+ig\delta(x-y) \int dx_1 dy_1\Big\{\tr_u[S(x-x_1) F^c_2\left(
\begin{array}{cc} x_1&y_1\\x'&y'\end{array} \right)
S(y_1-y)]-
$$
$$
-\gamma_5\mbox{\ttt}\tr_u[\gamma_5\mbox{\ttt}S(x-x_1) F^c_2\left(
\begin{array}{cc} x_1&y_1\\x'&y'\end{array} \right)
S(y_1-y)]\Big\}.
$$
Iterations of this equation reproduce the discrete algebraic
structure (colour, isotopic and Lorentz) of the inhomogeneous
term. This circumstance permits to write the general form of the
solution
\begin{equation}
F^c_2\left(
\begin{array}{cc} x&y\\x'&y'\end{array} \right)=
\delta(x-y)\delta(x'-y')\{1 \cdot 1
A_\sigma(x-x')+\gamma_5\mbox{\ttt}\cdot\gamma_5\mbox{\ttt}
A_\pi(x-x')\} \label{F2c}
\end{equation}
Here $A_\sigma$ is the scalar (sigma-meson) amplitude, $A_\pi$ is
the pseudoscalar (pion) amplitude. These amplitudes  are solutions
of  equations
$$
A_\sigma(x)=-ig\delta(x)+ig\int dx_1\tr
[S(x-x_1)S(x_1-x)]A_\sigma(x_1),
$$
$$
A_\pi(x)=ig\delta(x)-ig\int dx_1\tr
[\gamma_5S(x-x_1)\gamma_5S(x_1-x)]A_\pi(x_1).
$$

In  momentum space
\begin{equation}
A_\sigma(p)=\frac{1}{4n_c(4m^2-p^2)I_0(p)};\;\;
 A_\pi(p)=\frac{1}{4n_cp^2I_0(p)},
 \label{A_sigma}
\end{equation}
where
\begin{equation}
 I_0(p )=\int \frac{d^4q}{(2\pi)^4}\frac{1}
 {(m^2-(p+q)^2)(m^2-q^2)}. \label{I0}
\end{equation}
The integral in equation (\ref{I0}) is divergent and should be
dealt with by some regularization.

The equation for the NLO propagator $S^{(1)}$ is a system of
simple algebraic equations. Introducing the first-order mass
operator with formula  $\Sigma^{(1)}=S^{-1}\star S^{(1)}\star
S^{-1}$, we obtain for $\Sigma^{(1)}$ the following expression
\begin{equation}
\Sigma^{(1)}(x) =ig\delta(x)\tr[ S^{(1)}(0)]+
S(x)A_\sigma(x)+3\gamma_5S(x)\gamma_5 A_\pi(x). \label{Sigma}
\end{equation}

\section{Second-step equations. Solutions for four-particle
and three-particle
functions}

The second-step generating functional is
$$ G^{(2)}=
\biggl\{\frac{1}{4!}\mbox{Tr}\Big( S_4\ast\eta^4\Big)+
\frac{1}{3!}\mbox{Tr}\Big( S_3\ast\eta^3\Big) +
\frac{1}{2!}\mbox{Tr}\Big( S^{(1)}_2\ast\eta^2\Big) + \mbox{
Tr}\Big(S^{(2)}\ast\eta\Big)\biggr\}G^{(0)},
$$
i.e., second-step equations define four-particle function $S_4$,
three-particle function  $S_3$, and also NLO two-particle function
$S^{(1)}_2$ and NNLO propagator $S^{(2)}$. For these four
functions we have a system of four integral equations. All these
equations (and all  equations of following steps of the iteration
scheme) possess the common structure, which is similar to the
structure of equation (\ref{S2}):
\begin{equation}
S_n\left( \begin{array}{cc} x_1&y_1\\
\cdots&\cdots\\x_n&y_n\end{array} \right)=
S_{n}^0\left( \begin{array}{cc} x_1&y_1\\
\cdots&\cdots\\x_n&y_n\end{array} \right)+\label{Sn}
\end{equation}
$$
 +ig\int dx'_1\Big\{(S(x_1-x'_1)
S(x'_1-y_1)) \tr_u\Big[ S_n\left( \begin{array}{cc} x'_1&x'_1\\
\cdots&\cdots\\x_n&y_n\end{array} \right)\Big]-
$$
$$
 - (S(x_1-x'_1)\gamma_5\mbox{\ttt}
S(x'_1-y_1)) \tr_u\Big[ \gamma_5\mbox{\ttt}S_n\left(
\begin{array}{cc} x'_1&x'_1\\
\cdots&\cdots\\x_n&y_n\end{array} \right)\Big]\Big\}
$$
Difference becomes apparent in the structure of inhomogeneous
terms. The inhomogeneous term in the equation for $S_4$  is
\begin{equation}
S_{4}^0\left( \begin{array}{cc} x_1&y_1\\
x_2&y_2\\x_3&y_3\\x_4&y_4\end{array}
\right)=-\biggl\{S(x_1-y_2)S(x_2-y_1)S_2\left( \begin{array}{cc}
x_3&y_3\\x_4&y_4\end{array} \right)\biggr\}-\biggl\{
2\leftrightarrow 3\biggr\}-\biggl\{ 2\leftrightarrow 4\biggr\}
\label{S40}
\end{equation}
where $S_2$ is defined in preceding section, i.e. the equation for
$S_4$ does not include  another three second-step functions. The
inhomogeneous term in the equation for three-particle
function$S_3$ includes function $S_4$. This inhomogeneous term has
the form
\begin{equation}
S_{3}^0\left( \begin{array}{cc} x_1&y_1\\
x_2&y_2\\x_3&y_3\end{array}
\right)=-S(x_1-y_2)S(x_2-y_1)S^{(1)}(x_3-y_3)
-S(x_1-y_3)S(x_3-y_1)S^{(1)}(x_2-y_2)-\label{S30}
\end{equation}
$$
-S(x_1-y_2)S_{2}\left( \begin{array}{cc}
x_2&y_1\\x_3&y_3\end{array} \right)-S(x_1-y_3)S_{2}\left(
\begin{array}{cc} x_3&y_1\\x_2&y_2\end{array} \right)+
$$
$$
+ig\int dx'_1S(x_1-x'_1)\Bigl\{\tr_u[S_{4}
\left( \begin{array}{cc} x'_1&x'_1\\
x'_1&y_1\\x_2&y_2\\x_3&y_3\end{array}
\right)]-\gamma_5\mbox{\ttt}\tr_u[\gamma_5\mbox{\ttt}S_{4}
\left( \begin{array}{cc} x'_1&x'_1\\
x'_1&y_1\\x_2&y_2\\x_3&y_3\end{array} \right)]\Bigr\}
$$

Similarly, the inhomogeneous term in equation for $S_2^{(1)}$
includes function $S_3$, and the inhomogeneous term of the
equation for $S^{(2)}$ includes function  $S_2^{(1)}$.

 Due to such
structure of the system its solution should be started by equation
for four-particle function $S_4$, then should be solved the
equation for three-particle function
 $S_3$, etc.

 The equation for
four-particle function  $S_4$ with the inhomogeneous term
(\ref{S40}) has the simple solution (see also \cite{JaRo2})
\begin{equation}
S_{4}\left( \begin{array}{cc} x_1&y_1\\
x_2&y_2\\x_3&y_3\\x_4&y_4\end{array} \right)= S_2\left(
\begin{array}{cc} x_1&y_1\\x_2&y_2\end{array} \right)S_2\left(
\begin{array}{cc} x_3&y_3\\x_4&y_4\end{array} \right)+\biggl\{
2\leftrightarrow 3\biggr\}+\biggl\{ 2\leftrightarrow 4\biggr\}
\label{S4}
\end{equation}
This solution is disconnected, and it means the absence of
physical effects due to four-particle functions in the given order
of mean-field expansion. Particularly, the pion-pion scattering is
absent in the given order and will appear in the next order.

Solution (\ref{S4}) of the equation for four-particle function
gives us the closed equation   of type (\ref{Sn}) for
three-particle function $S_3$. The solution of this equation with
inhomogeneous term (\ref{S30}) can be obtained likewise  to
solving of equation (\ref{S2}) for the two-particle function. We
go to the amputated function
$$
F_3=S^{-1}\ast S^{-1}\ast S^{-1}\ast S_3\ast S^{-1}\ast S^{-1}\ast
S^{-1}
$$
and then separate  the functions with the algebraic structures,
which are reproduced after iterations. Connected part of the
amputated three-particle function possesses two-meson and
three-meson contributions. The explicit form of this function in
momentum space see in Appendix.

\section{Two-particle function and pion-decay constant}
The solution of the equation for three-particle function permits
 to close the equation for  NLO two-particle function $S_2^{(1)}$.
  This equation differs from equation (\ref{S2})
only by inhomogeneous term. The inhomogeneous term in  equation
for $S_2^{(1)}$ is
$$
S_{2}^{0(1)}\left( \begin{array}{cc} x&y\\
x'&y'\end{array} \right)=-S(x-y')S^{(1)}(x'-y)+
$$
$$
+ig\int dx_1S(x-x_1)\Bigl\{\tr_u[S_{3}
\left( \begin{array}{cc} x_1&x_1\\
x_1&y\\x'&y'\end{array}
\right)]-\gamma_5\mbox{\ttt}
\tr_u[\gamma_5\mbox{\ttt}S_{3}\left( \begin{array}{cc} x_1&x_1\\
x_1&y\\x'&y'\end{array} \right)]\Bigr\}
$$
The two-particle function of the second step enables to calculate
a correction to pion-decay constant $f_\pi$, which is one of the
basic strong-interaction parameters. The pion-decay constant is
defined by the relation
$$
if_\pi\delta^{bb'}P_\mu=<0\vert J_{\mu 5}^b\vert P, b'>,
$$
where $\vert P, b'>$ is the pion state with momentum $P$ and
isospin $b$, and $J_{\mu
5}^b=\bar{\psi}\gamma_\mu\gamma_5\frac{\tau^b}{2}\psi$ is the
axial current. If the two-particle function has a pole term
$S_2^{pole}$, which corresponds to the pion,  then, taking into
account  these definitions, this pole term is connected with the
pion-decay constant by relation
\begin{equation}
\tilde{\delta}(P-P')f^2_\pi=i\int dx dx'e^{i(Px-P'x')} \tr_{u, d}
[\gamma_\mu\gamma_5\frac{\mbox{\ttt}}{2}
\cdot\gamma_\mu\gamma_5\frac{\mbox{\ttt}}{2} S_2^{pole}\left(
\begin{array}{cc} x&x\\x'&x'\end{array}\right)] \label{fpi}
\end{equation}
Here $\tr_{u, d}$ denotes the traces over up and down lines of
function $S_2^{pole}$.

In the leading order we obtain from (\ref{fpi}) the well-known
expression for the pion-decay constant (see, for example,
\cite{kle}):
\begin{equation}
(f^{(0)}_\pi)^2=-4i n_c m^2I_0(0), \label{fpi0}
\end{equation}
 where $I_0$ is defined by formula (\ref{I0}). For a
 regularization with four-dimensional cutoff \\ $
I_0(0)= \frac{i}{(4\pi)^2}\Bigl[\log\frac{\Lambda^2+m^2}{m^2}-
\frac{\Lambda^2}{\Lambda^2+m^2}\Bigr],$ where  $\Lambda$ is the
cutoff parameter.

In the next-to-leading order formula (\ref{fpi}) defines
correction $(f^{(1)}_\pi)^2$ to expression (\ref{fpi0}). Surely,
to calculate this correction it is no need in a complete solution
of the NLO two-particle equation. In correspondence with
(\ref{fpi}) it is quite enough to calculate the pion-pole part
only.

  The results of calculation of ratio
$r_f=\frac{(f^{(1)}_\pi)^2}{(f^{(0)}_\pi)^2}$ in the scheme with
four-dimensional cutoff are shown  in Table  for three sets of
model parameters. These sets are correspond to different values of
quark condensate $c=(\frac{1}{2}<\bar{\psi}\psi>)^{1/3}$ at
physical value of pion-decay constant $f_\pi=93\;MeV$.

\vspace{1cm}

\begin{center}
\begin{tabular}{|l|l|l|l|l|}\hline
 $c$ (GeV)& $m$ (GeV) & $\Lambda$ (GeV)& $\kappa=3g\Lambda^2/2\pi^2$ &
  $r_f=(f^{(1)}_\pi)^2/(f^{(0)}_\pi)^2$\\
 \hline
\  -0.21  & \ 0.42  & \  0.73  & \   1.87  & \ 0.14\\
\  -0.23  & \ 0.28 &  \  0.87 & \   1.33 & \ 0.20\\
\  -0.25 &  \ 0.24  &  \ 1.03   & \  1.19 & \ 0.28\\

 \hline

\end{tabular}
\vspace{1cm}

{\small Table. Values of ratio
$r_f=(f^{(1)}_\pi)^2/(f^{(0)}_\pi)^2$ in scheme with
four-dimensional cutoff for different sets of values of chiral
quark condensate
 $c$, quark mass  $m$, regularization parameter  $\Lambda$
 and dimensionless coupling   $\kappa$.}
\end{center}

\vspace{1cm}

Apparently the correction to value of  the pion-decay constant
increases from 14\% to 28\%  when  the absolute value of the quark
condensate   increases from $(0.21\; GeV)^3$ to $(0.25\; GeV)^3$.
These values of the correction to pion-decay constant are not far
from results of calculations  in works
\cite{Nikolov}-\cite{Oertel}, though the method of calculations is
different. This principal correspondence seems to be  a reflection
of some stability of NJL model with regard to quantum corrections.

\section*{Acknowledgments}

Author  is grateful to Dr. R.G. Jafarov for useful discussions.

\newpage

\section*{Appendix}

The connected part of the three-particle funcion in the momentum
space is
$$
F_3^c=F_3^{two-meson}+F_3^{three-meson}
$$
$$
F_3^{two-meson}= -(S(p+P')+S(p+P''))\cdot \mbox{A}_\sigma(P')\cdot
\mbox{A}_\sigma(P'')-
$$
$$-\mbox{A}_\sigma(P)\cdot(S(p'+P'')+S(p'+P))\cdot
\mbox{A}_\sigma(P'')-
$$
$$
-\mbox{A}_\sigma(P)\cdot \mbox{A}_\sigma(P')\cdot
(S(p''+P)+S(p''+P')) +
$$
$$
+(\gamma_5S(p+P')\gamma_5+\gamma_5S(p+P'')\gamma_5)\cdot {\bf
A}_\pi(P')\cdot {\bf A}_\pi(P'') +
$$
$$+\mbox{A}_\sigma(P)\cdot(\gamma_5S(p'+P'')
+S(p'+P)\gamma_5){\bf\mbox{\ttt}}\cdot
{\bf A}_\pi(P'')+
$$
$$+\mbox{A}_\sigma(P)\cdot {\bf A}_\pi(P')\cdot {\bf\mbox{\ttt}}
(S(p''+P)\gamma_5+\gamma_5S(p''+P')) +
$$
$$+(S(p+P')\gamma_5+\gamma_5S(p+P'')){\bf\mbox{\ttt}}\cdot
\mbox{A}_\sigma(P')\cdot {\bf A}_\pi(P'')+
$$
$$+{\bf A}_\pi(P)\cdot(\gamma_5S(p'+P'')\gamma_5
+\gamma_5S(p'+P)\gamma_5)\cdot
{\bf A}_\pi(P'')+
$$
$$+{\bf A}_\pi(P)\cdot
\mbox{A}_\sigma(P')\cdot {\bf\mbox{\ttt}}
(\gamma_5S(p''+P)+S(p''+P')\gamma_5) +
$$
$$
+(\gamma_5S(p+P')+S(p+P'')\gamma_5){\bf\mbox{\ttt}}\cdot {\bf
A}_\pi(P')\cdot  \mbox{A}_\sigma(P'') +
$$
$$
+{\bf
A}_\pi(P)\cdot{\bf\mbox{\ttt}}(S(p'+P'')\gamma_5+\gamma_5S(p'+P))\cdot
\mbox{A}_\sigma(P'')+
$$
$$
+{\bf A}_\pi(P)\cdot {\bf A}_\pi(P')\cdot
(\gamma_5S(p''+P)\gamma_5+\gamma_5S(p''+P')\gamma_5) +
$$
$$
+ i
(\gamma_5S(p+P')\gamma_5-\gamma_5S(p+P'')\gamma_5){\bf\mbox{\ttt}}\cdot
 [ {\bf A}_\pi(P')\times {\bf A}_\pi(P'')]+
$$
$$
+i {\bf A}_\pi(P)\cdot  [(\gamma_5S(p'+P'')\gamma_5
-\gamma_5S(p'+P)\gamma_5){\bf\mbox{\ttt}} \times {\bf A}_\pi(P'')]
+
$$
$$
+i {\bf A}_\pi(P)\cdot [ {\bf A}_\pi(P')\times {\bf\mbox{\ttt}}
(\gamma_5S(p''+P)\gamma_5 -\gamma_5S(p''+P')\gamma_5)].
$$
$$
F_3^{three-meson}=2n_c\mbox{A}_\sigma(P)\cdot
\mbox{A}_\sigma(P')\cdot \mbox{A}_\sigma(P'')[\Delta_{sss}(P;
P+P')+ \Delta_{sss}(P; P+P'')]-
$$
$$
-2n_c\mbox{A}_\sigma(P)\cdot {\bf A}_\pi(P')\cdot {\bf
A}_\pi(P'')[\Delta_{spp}(P; P+P')+ \Delta_{spp}(P; P+P'')]-
$$
$$
-2n_c{\bf A}_\pi(P)\cdot {\bf A}_\pi(P')\cdot
\mbox{A}_\sigma(P'')[\Delta_{spp}(P''; P'+P'')+\Delta_{spp}(P'';
P+P'')]-
$$
$$
-2n_c{\bf A}_\pi(P)\cdot \mbox{A}_\sigma(P')\cdot {\bf
A}_\pi(P'')[\Delta_{spp}(P'; P'+P'')+ \Delta_{spp}(P'; P+P')].
$$
Here $ p= p_{x_1},\; P= p_{x_1}+p_{y_1},\; p'= p_{x_2},\; P'=
p_{x_2}+p_{y_2},\;p''=p_{x_3},\; P''= p_{x_3}+p_{y_3}$, where
$p_{x_i}, p_{y_i}$ are quark momenta    and $P+P'+P''=0$. We also
use the notations
 $
\mbox{A}_\sigma=1\circ A_\sigma $ for  the scalar-singlet
function,
 $ {\bf A}_\pi=
\gamma_5{\bf\mbox{\ttt}}\circ A_\pi $ for the
pseudoscalar-isovector function (see (\ref{A_sigma})) and
$$
\Delta_{sss}(P; Q)=\int d\tilde{q}\tr[S(P+q)S(Q+q)S(q)],\;
\Delta_{spp}(P; Q)=\int
d\tilde{q}\tr[S(P+q)\gamma_5S(Q+q)\gamma_5S(q)]
$$
for  quark triangles. (In  last formulae the traces are taken over
Lorentz spinor indices only.)

\end{document}